\documentclass{aa}
\usepackage{booktabs}
\usepackage{mathtools}
\usepackage{natbib}
\usepackage{graphicx}
\usepackage{pgf}
\usepackage{txfonts}

\usepackage{textcomp}
\usepackage{gensymb} 
\usepackage{amsmath}
\usepackage{tabularx}

\usepackage{threeparttable}
\usepackage{mathtools}

\usepackage{soul}
\definecolor{lightgreen}{HTML}{B7F774}
\definecolor{lightred}{HTML}{FF6666}
\definecolor{lightorange}{HTML}{FE9A2E}
\sethlcolor{lightgreen}

\usepackage[normalem]{ulem}

\newcommand{\unit}[1]{\ensuremath{\, \mathrm{#1}}}

\begin{document} 

\title{Detection of multiple velocity components\\ in partially overlapping
emitting regions}

\author{F. Mertens\inst{1}
        \and
        A.~P. Lobanov\inst{1,2}
        }

\institute{Max-Planck-Institut f\"ur Radioastronomie,
          Auf dem H\"ugel 69, 53121 Bonn, Germany
          \and
          Institut f\"ur Experimentalphysik, Universit\"at Hamburg, 
          Luruper Chaussee 149, 22761 Hamburg, Germany
          }

\date{}

\abstract
{Velocity measurements made from multiple-epoch astronomical images of evolving objects
  with optically thin continuum emission (e.g. as relativistic jets or
  expanding supernova shells) may be confused as a result of the overlap
  of semi-transparent features moving at different speeds.}
{Multi-scale wavelet decomposition can be effectively applied  to
  identify and track such overlapping features, provided that their
  respective structural responses can be separated over the spatial
  scales used for the decomposition.}
{We developed a new method that combines the stacked
  cross-correlation with the wavelet-based image
segmentation and evaluation (WISE) technique of decomposition of
  two-dimensional structures, to separate and track dominant spatial responses of overlapping
  evolving features.}
{The method is tested on a set of simulated images of a stratified
  relativistic jet, demonstrating the robust detection of both the faster
  spine and the slower sheath speeds. The method is applied to
  mutliple-epoch images from the MOJAVE survey, revealing two
  different superluminal streams inside the jet in 3C\,273 and the acceleration of
  the flow in 3C\,120.}
{The method can be applied to densely monitored objects with composite
  structural evolution such as the parsec-scale jet in M\,87 or
  heavily resolved expanding supernova shells.}


\maketitle

\section{Introduction}

Images of various astronomical objects, such as relativistic jets or
supernova shells, display dynamically evolving, optically thin
continuum emission, which is likely to feature partially overlapping
semi-transparent emitting regions with different kinematic and
structural properties.  In this situation, analysis of structural
evolution of the emitting material often becomes
difficult and inconclusive.  The ensuing ambiguity of interpretation
of the structural changes can be resolved in some cases with the help
of spectroscopic measurements~\citep[e.g.][]{guillard_turbulent_2012}. 

However, the application of spectroscopy is limited to objects with
substantial line emission that can be kinematically associated with
the same material that emits the continuum emission. No such
association can be made in objects dominated by the continuum: for
instance in objects in which the continuum is generated in a
relativistic plasma. Radio emission from extragalactic jets presents
a strong example of such a setting. Optically thin synchrotron emission
and prominent transverse stratification of the flow are expected to be
found in jets as a direct consequence of the jet formation mechanism
\citep{sol_twoflow_1989,gracia_magnetic_2005,komissarov_magnetic_2007}
or as a result of shocks \citep{lobanov_spectral_1999,lobanov_dualfrequency_2006} and
instability development
\citep{lobanov_vsop_1998,lobanov_cosmic_2001,perucho_role_2006,perucho_physical_2007a,perucho_s5_2012}.
Observational measurements of the flow stratification in relativistic
jets are critically needed to understand the physical processes
governing the dynamics of the emitting plasma
\citep[cf.][]{lobanov_a_2003,hardee_modeling_2005,walker_vlba_2008,mertens_longitudinal_2014}.

Wavelet decomposition has been applied in a number of works to
analyse the structure of extended objects in astronomical images \citep[cf.][]
{starck_astronomical_2006}. The wavelet-based image
segmentation and evaluation (WISE) algorithm developed for determining
two-dimensional velocity fields \citep{mertens_waveletbased_2015} can
also be used to differentiate between structural components dominating
at different spatial scales. The principal feasibility for WISE to
detect multiple velocity components has already been demonstrated
\citep{mertens_waveletbased_2015}. In this paper, we extend this
capability to partially overlapping optically thin regions, by
combining the WISE approach with stacked-cross-correlation (SCC)
\citep[cf.][]{fuhrmann_detection_2014}.

The basic approach for applying SCC to
the detection of multiple velocity components is introduced in
Sect.~\ref{sc:scc} and tested on simulated images containing two
overlapping optically thin streams. In Sect.~\ref{sc:app}, the SCC
method is applied to MOJAVE data on the jets in 3C\,120 and 3C\,273,
revealing the complexity of the kinematic parameters of these
outflows. Further potential applications of the method are discussed
in Sect.~\ref{sc:discuss}.

Throughout this paper, we adopt a cosmology of 
$H_0 = 71\ \unit{km\ s^{-1}\ Mpc^{-1}}$, $\Omega_M = 0.27$ and $\Omega_\Lambda =
0.73$~\citep{lister_mojave_2013}.

\section{Stacked cross-correlation for detection of multiple velocity components}
\label{sc:scc}

The WISE algorithm combines segmented wavelet decomposition (SWD) and multi-scale
cross-correlation (MCC) to provide robust identification and tracking of
structural patterns in astronomical images. The SWD algorithm provides a
structural representation of astronomical images with exceptional sensitivity
for identifying compact and marginally resolved features as well as large scale
structural patterns. The SWD decomposition delivers a set of two-dimensional
significant structural patterns (SSP), which are identified at each scale of the
wavelet decomposition. Tracking of these SSP detected in multiple-epoch images
is performed with a MCC algorithm. It combines structural information on
different scales of the wavelet decomposition and provides a robust and reliable
cross-identification of related SSP.  \cite{mertens_waveletbased_2015} provide a
full description of the method.

Initial results of application of WISE have indicated that the method can
discriminate overlapping features with different velocities
\citep{mertens_waveletbased_2015}. The successful detection of multiple velocity
components has however remained challenging and prone to an increasing failure
rate at low signal-to-noise ratios. This deficiency can be alleviated with the
use of stacked cross-correlation.

The displacement of a single SSP can be inferred from the location of the
maximum of the cross-correlation obtained for this SSP between two given epochs.
In optically thin objects with  complex and stratified three-dimensional
structures, the resulting cross-correlation map may contain several peaks
corresponding to several potential displacements. The 
MCC procedure resolves the
potential ambiguity of interpretation of these peaks by combining the
displacements found at different scales of the wavelet decomposition and
determining the group motion of causally connected SSP.

\begin{figure}
    \centering
    \includegraphics{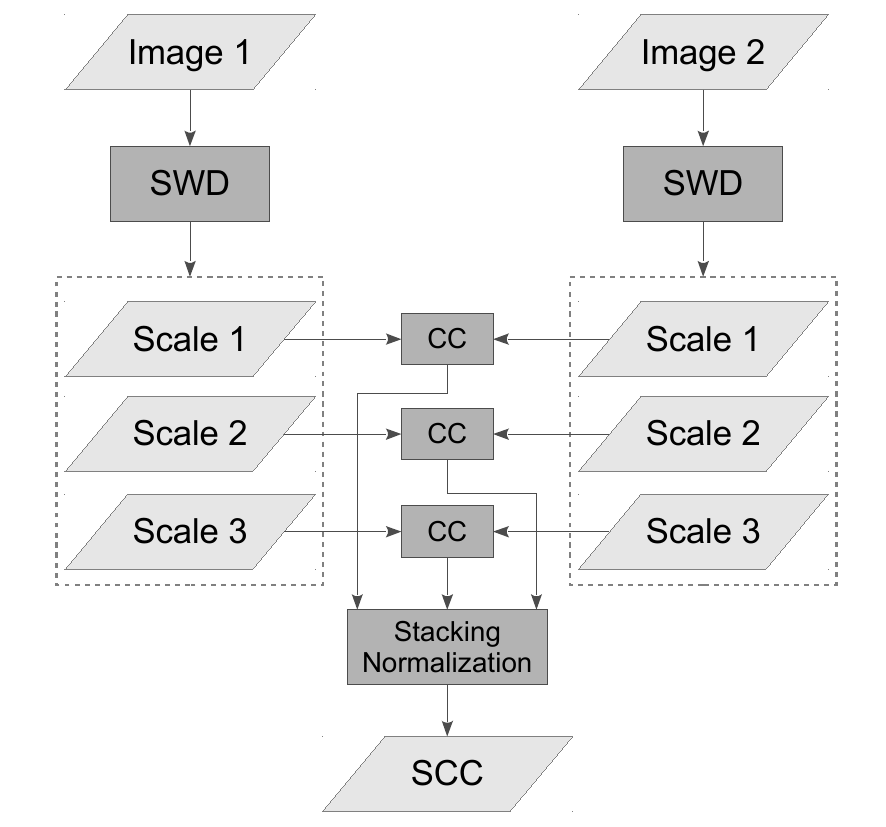}
    \caption[Stacked cross-correlation analysis of simulated of jet.]
    {Simplified scheme of the stacked cross-correlation (SCC) algorithm for
determining multiple velocity components. The SCC combines the SSP identified at
different scales of the SWD and (in case of more than two images analysed)
different pairs of epochs. }
  \label{fig:scc_scheme}
\end{figure}

A different approach can be adopted if the stratification is
homogeneous over an extended region (e.g. one is dealing with two
large-scale flows with different but constant velocities) and/or if
several observations of the same steady stratified flow are available. In
this case, it is possible to detect two or more main velocity
components, corresponding to different layers of the flow consistently
over an extended region and across several scales of the SWD, and the
cross-correlation response of several SSP may be joined or
stacked. The basic outline of the resulting stacking cross-correlation
algorithm is illustrated in Fig.~\ref{fig:scc_scheme}.

The SWD provides scale-dependent models (SDM) of the target image by identifying
SSP at each wavelet scale. The SCC algorithm combines the information contained
in the scale-dependent models of two images (reference and target image) in the
following way:

\begin{itemize}

\item[1.]~The SWD is used to identify sets of SSP in both images at
  different scales of the wavelet decomposition.

\item[2.]~The SSPs of the reference image are cross-correlated with
  the target image.

\item[3.]~The cross-correlation results for all SSP, all
  scales, and (in case of more than two images analysed) different
  pairs of epochs are stacked together and normalized.

\end{itemize}

\noindent
This procedure yields a two-dimensional distribution of the cumulative
correlation coefficients, with peaks corresponding to different
velocity components. The significance and uncertainty of each of these
peaks can be evaluated using Monte Carlo simulations.

\subsection{Segmented wavelet decomposition}
\label{sect:swd}

The SWD used at the first step of the SCC algorithm comprises the following
steps to describe an image structure with a set of
SSP~\citep{mertens_waveletbased_2015}:

\begin{itemize}
\item[1.]~A wavelet transform is performed on an image $I$  by
decomposing the image into a set of $J$ sub-bands (scales), $w_{j}$,
  and estimating the residual image noise (variable across the image).

\item[2.]~At each sub-band, statistically significant wavelet
  coefficients are extracted from the decomposition by thresholding
  them against the image noise.

\item[3.]~The significant coefficients are examined for local maxima,
  and a subset of the local maxima satisfying composite detection
  criteria is identified. This subset defines the locations of SSP in
  the image.

\item[4.]~Two-dimensional boundaries of the SSP are defined 
  by the watershed segmentation using the feature
  locations as initial markers.

\end{itemize} 

\noindent
The resulting SDM representation of an image at the
scale $j$ is derived as a group of SSP as
\begin{equation}
 S_{j} = \{s_{j, i}: i=1,...,N_\mathrm{SSP,j}\}\,.
\end{equation}

\noindent
The combination of all SDMs provides a structure representation that is
sensitive to compact and marginally resolved features and to structural
patterns that are much larger than the full width at half maximum (FWHM) of the
instrumental point spread function (PSF) in the image.

A well approximated estimate of the uncertainty on the SSP position can be
derived
following \cite{fomalont_e_1999}:
\begin{equation}
\label{eq:feature_pos_error}
\sigma_x = \frac{b_x}{\sqrt{2}\ \mathrm{SNR}},\ \sigma_y = \frac{b_y}{\sqrt{2}\ \mathrm{SNR}} 
,\end{equation}

\noindent
where $b_x$ and $b_y$ are the beam size along the x and y coordinate, respectively,
assuming an elliptical PSF, which is approximately the case for application to
VLBA maps, and $\mathrm{SNR }$ is the signal-to-noise ratio of the SSP.

\subsection{Stacked cross-correlation}
\label{sect:scc}

To detect structural changes between the reference and target images, each
single SSP $s_{j, i}$ identified on a scale $j$ of the reference image is 
cross-correlated with the target image. For this purpose, a zero-mean cross-correlation, insensitive to both the image intensity offset and scale change, is
used and is defined as
follows~\citep{giachetti_matching_2000,mertens_waveletbased_2015}:
\begin{equation}
C_\mathrm{ZNCC}(a, b) = \frac{\sum \overline{a_i} \overline{b_i
}}{\sqrt{\sum \overline{a_i }^2 \sum \overline{b_i }^2}}\,,\end{equation}
where $a$ and $b$ the two images to cross-correlate.

\noindent The cross-correlation $\gamma_{i, j, t}$ between an SSP $i$ detected
at scale $j$ of reference epoch $t$, and target epoch $t + 1$ is then written as
\begin{equation} \gamma_{i, j, t} =
C_\mathrm{ZNCC}(s^{t}_{j, i}, w_j^{t+1}) .\end{equation}

In order to reduce the adverse impact of the image noise on the calculation, we
consider here only prominent peaks $p$ above a certain threshold $\kappa$
(typically, 0.6--0.8) and model them as two-dimensional Gaussian shapes $g(p,
\sigma)$ with the width $\sigma$ corresponding to the error of the displacement
(typically a fraction of the beam size). Additionally, if the total set of
analysed images consists of multiple epochs with inhomogeneous time intervals
between the successive epochs, an additional linear transformation $\mathbf
{V_t}$ that converts displacement to velocity needs to be applied so that all
cross-correlation results can be stacked together. The resulting modified cross-correlation coefficient is then defined as 
\begin{equation} \tilde{\gamma}_{i,
j, t} = \sum_{p \in P}{\gamma_{i, j, t}(p) g(\mathbf{V_t}p, \mathbf{V_t}\sigma)}
,\end{equation} 

\noindent where $P = \{p\}$ is the group of local maxima of $\gamma_{i, j, t}$,
which are above threshold $\kappa$.

Finally, combining all the SSP identified at different scales of the SWD and
across different
epochs, we define the stack cross-correlation as
\begin{equation}
C_\mathrm{SCC} = \frac{\sum_{t}\sum_{j}\sum_{i}\tilde{\gamma}_{i, j, t}}
{\sum_{t}\sum_{j}N^{t}_{\mathrm{SSP,j}}}
,\end{equation}

\noindent where $N^{t}_{\mathrm{SSP,j}}$ are the number of SSP identified at scale
$j$ of the SWD in the image of epoch $t$. The complete procedure yields a 
two-dimensional representation of the global correlation between all images
analysed with peaks corresponding to the main displacement/velocity components.

\subsection{Significance of detected velocity components}
\label{sect:scc_significance}

To ensure robust identification of different velocity components, we need to
determine the significance of individual maxima found using the stacked cross-correlation. If the cross-correlation is obtained from multiple epoch data,
bootstrapping can be applied to determine the probability for a given
correlation coefficient to  result from an inadequate sampling. The
bootstrapping consists of computing test SCCs for a large number of trials
(typically 1000) in which the individual epochs are randomly shuffled. The
confidence interval (CI) for each velocity components can then be calculated
from the mean and standard deviation of the test SCCs. For the velocities
corresponding to small displacements (typically, below 1/2 the beam size), the
mean coefficient of the shuffled sets are overestimated and corrected CI
 have to be computed in this case by adding a small shift (1/5 the beam
size) to the images.

\subsection{Uncertainty on the velocity components}
\label{sect:scc_error}

\begin{figure*}
    \centering
    \includegraphics{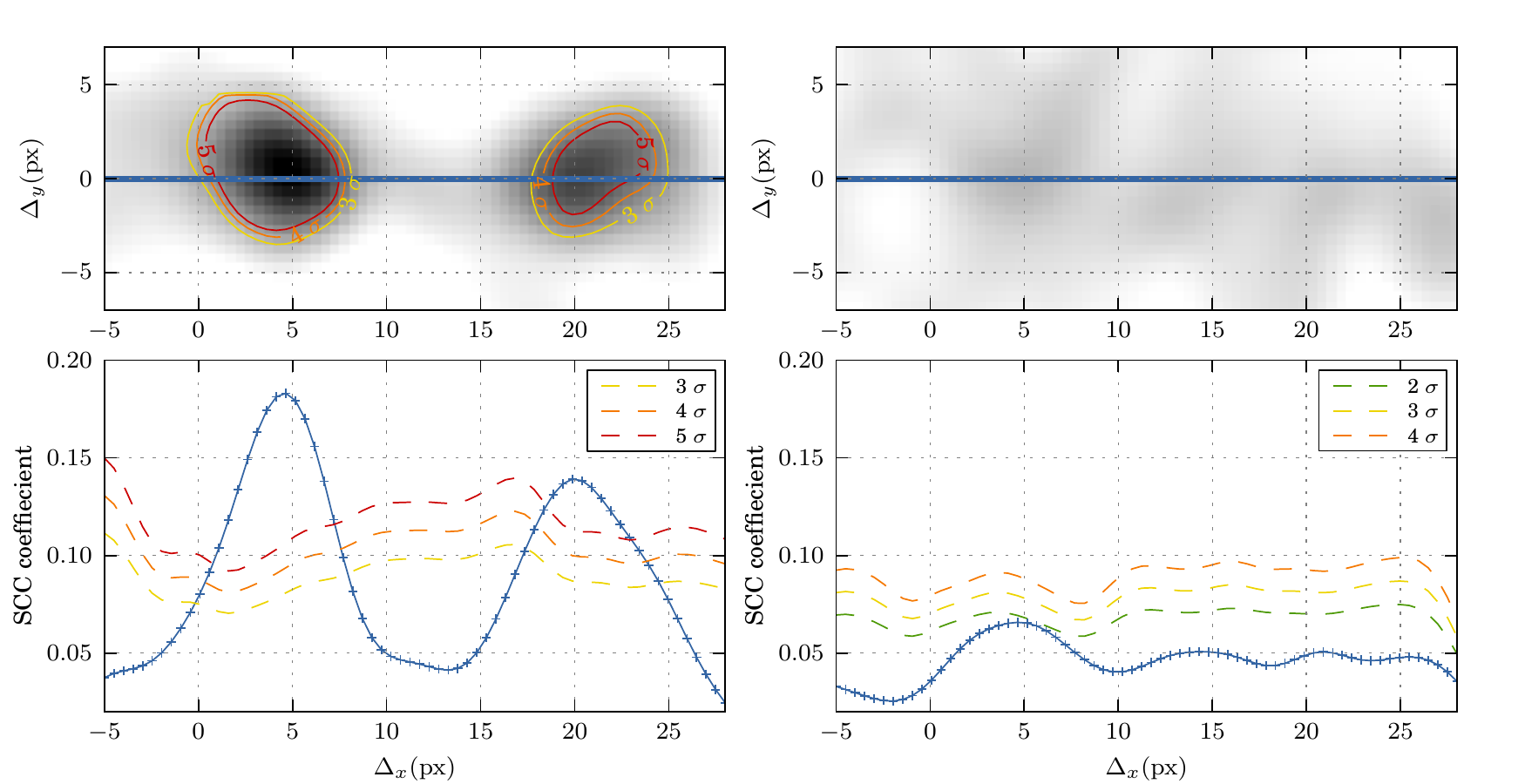}
    \caption[Stacked cross-correlation analysis of simulated jet.]
    {Stacked cross-correlation analysis of simulated images of a spine-sheath
      (``two-fluid'') jet with two constant velocity components (left)
      and a jet with random velocity field (right). For each test, a total of
      ten epochs are simulated. The
      resulting cross-correlation map is shown in greyscale in the
      respective top panel. The bottom panels shows the derived
      significance of the velocity components obtained. The
      significance is measured along the slice corresponding to the
      zero transverse speed. For the
      stratified spine-sheath flow on the left, the SCC successfully
      identifies the velocities of the two layers with a significance
      of $\ge 5\,\sigma$. For the jet on the right with the random
      velocity field, no peak with a significance greater than
      $2\,\sigma$ is found.  }
    \label{fig:gncc_simu_full}
\end{figure*}

An estimate of the uncertainty of a significant velocity component can
be obtained from Monte Carlo simulations. A large number of test SCCs are
again computed, this time with all the SSP randomly shifted by an
amount derived from a normal distribution with a standard deviation
set by the uncertainty on the SSP location (see
sect.~\ref{sect:swd}). 
%The position of a velocity component formally
%identified by the SCC will be shifted in the test SCC, and the standard
%deviation of this shift correspond to the uncertainty on the velocity
%component. 
It should be noted that for this measure to be valid, the
distribution of shifts must be Gaussian with the
mean corresponding to the velocity component identified with the SCC.

\subsection{Testing the SCC method}
\label{sc:tests}

To test the performance of the SCC method, two sets of simulated images of a
compact jet are prepared, following the prescription developed for testing the
WISE method \citep{mertens_waveletbased_2015}. The first set features a 
``two-fluid'' jet with a faster inner spine and a slower outer sheath. In the second
set, random velocity distribution is simulated. Images at multiple epochs are
generated for each set with structural displacements introduced accordingly.
The SCC is then computed for all consecutive image pairs. Examples of the
resulting cross-correlation map is shown in the top panels of
Fig.~\ref{fig:gncc_simu_full}. Peaks in the map are then located by searching
for local maxima.  A 2D Gaussian is fitted to a small box with dimensions equal to the beam size
around each local maximum found  to obtain sub-pixel precision on the peak location. The significance of each peak and its
associated velocity error are determined following the procedures introduced in
sect~\ref{sect:scc_significance} and sect~\ref{sect:scc_error}.

For the two-fluid scenario, the jet is simulated with one group of features
propagating with $\Delta x_1 = 5$\,px, $\Delta y_1 = 0$\,px between each pair of
images, and a second group of features evolving with $\Delta x_2 = 20$\,px,
$\Delta y_2 = 0$\,px between each pair of images (with $x$ and $y$ describing
the longitudinal and transverse dimensions, respectively). Noise is added to the
simulated displacements with $\sigma_x = \sigma_y = 2$\,px or 1/5 of the beam
size, which is set to 10\,px in all simulated images. Thus the faster jet
component moves by twice the beam size and the slower jet component moves by half the
beam size between each two consecutive epochs

A total of ten epochs is generated and SCC is computed in a region that
extends over 10 beam sizes longitudinally. The SCC includes the scales 1, 2, and
3 of the SWD decomposition, producing a total of 264 SSP identified in the jet
over all epochs. Results of this test are shown in the left panels of
Fig.~\ref{fig:gncc_simu_full}. The SCC velocity map obtained in simulation
shows two prominent peaks at velocities corresponding to the two simulated
displacements. No other peaks are found. For the first peak, a displacement of
$\Delta x_1 = 4.5 \pm 0.4$\,px and a significance of $11.4\,\sigma$ is found.
For the second peak, we measure $\Delta x_1 = 20.5 \pm 0.8$\,px with a
significance of $6.7\,\sigma$. As a cross check, we confirm that for both peaks
the distribution of displacements found in the test SCC used to measure the
uncertainties is Gaussian with mean value corresponding to the formal
displacements. These results are in excellent agreement with the simulated
displacements. The uncertainties of the displacement detection are improved by a
factor of 3 to 4 in comparison with a similar test performed using the WISE
analysis \citep[Sect.~5.3.2 in][]{mertens_waveletbased_2015}.

In the second simulated set, jet images are produced without any
significant regular velocity component. To achieve this, we simulate a
single group of features with random displacements drawn from a
uniform distribution $0\, \mathrm{px} \le \Delta_x \le 40\,
\mathrm{px}$ and $-5\, \mathrm{px} \le \Delta_y \le 5\,
\mathrm{px}$ between each pair of images. A similar procedure as described for
the first test is
then applied to determine the locations and the significance
of the correlation peaks. The results are shown in the right panels of
Fig.~\ref{fig:gncc_simu_full}. No peaks are found with significance
exceeding $2\,\sigma$, which is compatible with the simulated random
displacements.
These tests demonstrate the robustness of the SCC method for
identifying multiple velocity components manifested by partially
overlapping regions of the flow.

\section{Applications to astronomical images}
\label{sc:app}

After testing the SCC on simulated data, we have applied it to several
image sequences obtained as part of the MOJAVE\footnote{Monitoring of
  jets in active galactic nuclei with VLBA Experiments} long-term
monitoring program of extragalactic jets with Very Large Baseline
Interferometry (VLBI) observations \citep[][and references
therein]{lister_mojave_2013}. The particular focus of this analysis
was made on prominent radio jets in the quasar 3C\,273 and the radio
galaxy 3C\,120.

\subsection{Stratification in the jet of 3C\,273}

The jet in 3C\,273 is transversely resolved in the MOJAVE
images. Stratification of the flow has been previously suggested for
this jet \citep{lobanov_cosmic_2001,perucho_role_2006} and a large range of
apparent speeds has been reported at similar core separation based on
the analysis of the MOJAVE data
\citep{lister_mojave_2013,savolainen_multifrequency_2006}.

The SCC has been performed on the entire MOJAVE data for 3C\,273
comprising 69 images obtained in the 1996--2010 period. The average
beam (PSF) is $0.5\,\mathrm{mas}\times 1.0$\,mas in these images.  The WISE
decomposition was performed using four wavelet scales with the finest
scale of $0.2$\,mas (scale 1, corresponding to one fifth of the beam). 

\begin{figure}
    \centering
    \includegraphics{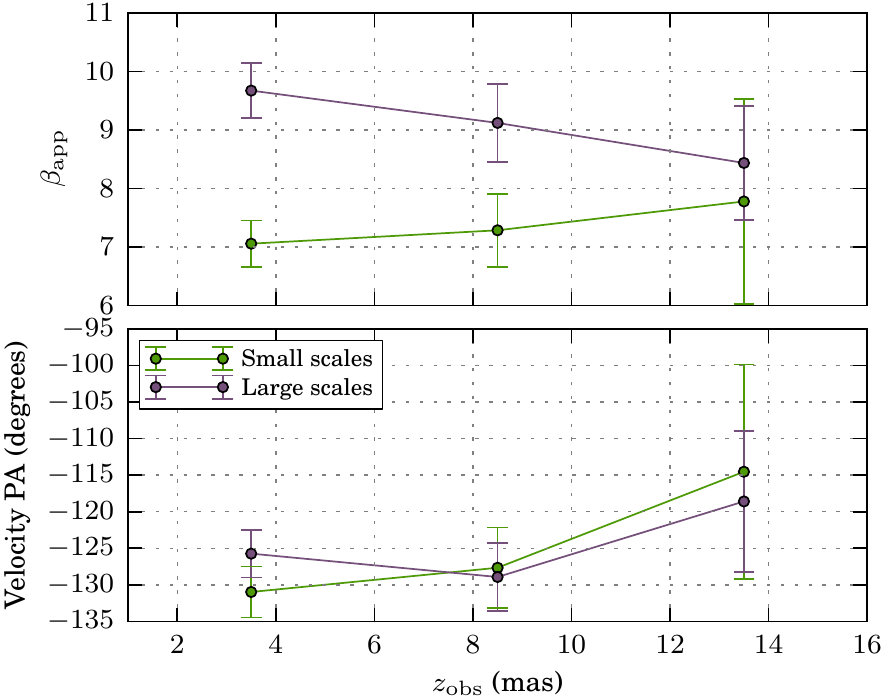}
    \caption
    {SCC analysis of the jet in 3C 273. The analysis is performed in three
      different regions in the jet: between 1 and 6 mas, 6 and 11 mas,
      11 and 16 mas. The SSP detected at larger scales (scales 3
      and 4, corresponding to 0.8\,and 1.6\,mas) of the SWD
      decomposition are
      found to be on average
      faster than SSP detected at smaller scale (scales 1 and
      2, corresponding to 0.2\,and 0.4\,mas).}
    \label{fig:3c273_stratification}
\end{figure}

\begin{figure}
    \centering
    \includegraphics{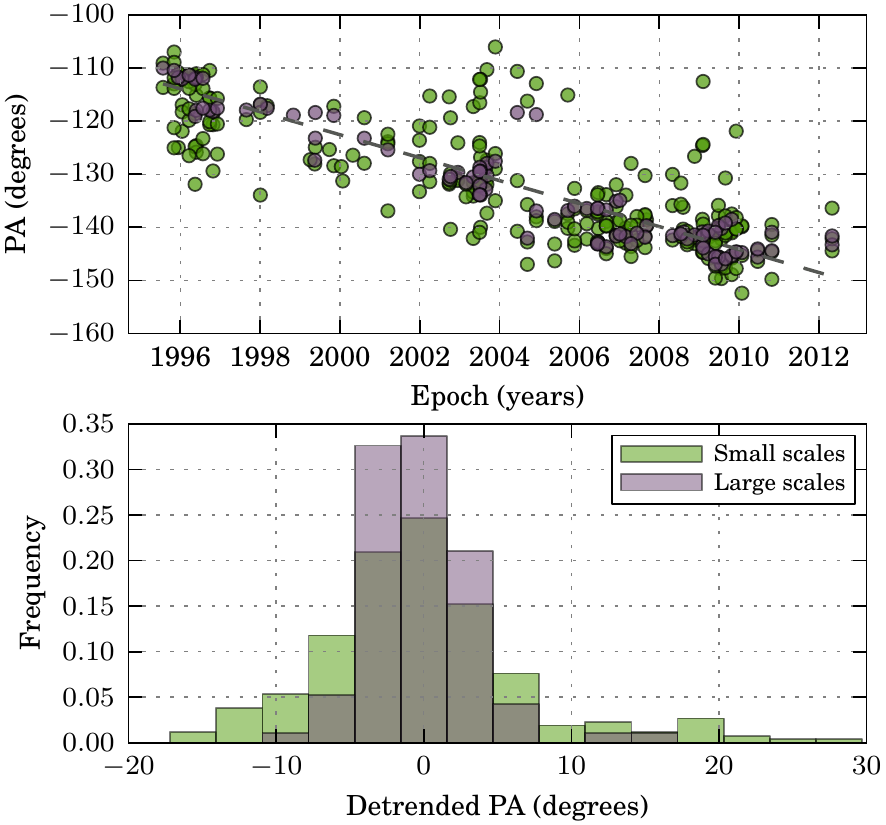}
    \caption
    {Analysis of the position angle of the SSP detected in the inner 6\,mas
region of the jet in 3C\,273 on small scales (scales 1 and 2) compared to that
detected at larger scales (scales 3 and 4). The comparison indicates that
the slower SSP detected on scales 1 and 2 are predominantly located in the outer
part of the jet, while the faster SSP detected on scales 3 and 4 are concentrated
towards the centre of the flow.}
    \label{fig:3c273_pa_frequency_1_6mas}
\end{figure}

The SCC results are shown in Fig.~\ref{fig:3c273_stratification}, indicating a
systematically higher speed of $\beta_\mathrm{app} \sim 9$\,c, detected at the
two larger scales (scale 3 and 4) compared to the slower apparent speed of
$\beta_\mathrm{app} \sim 7.5$\,c measured at the smaller scales (scale 1 and 2).
The position angle distribution of the SSP identified at the larger and smaller
scales (Fig.~\ref{fig:3c273_pa_frequency_1_6mas}) indicates that the faster
moving SSP are concentrated towards the central spine of the flow, while the
slower moving SSP tend to be found in the outer layers of the flow. This may be
indicative of  jet stratification, which results from either true transverse
velocity distribution in the flow or produced by fast shocks or disturbances
traveling through a slower underlying flow.

We note that the average apparent velocity of all components identified by the
MOJAVE team in this source is $\beta_\mathrm{app} \sim 8$\,c (with a standard
deviation of $\sim 3$\,c), which is similar to the average of the large- and
small-scale velocity obtained from the SCC analysis. The SCC was thus able to
single out in this source the two major superluminal streams responsible for the
large velocity discrepancy found by the MOJAVE team and to assign them to
structure of different scales.

\subsection{Acceleration in the jet of 3C\,120}

The analysis of the jet in 3C\,120 comprises the total of 87 MOJAVE images
covering the period of 1996--2010. The average beam of the MOJAVE images is $0.5
\,\mathrm{mas}\times 1.2$\,mas. The SWD decomposition was performed using four
wavelet scales
ranging between 0.2\,mas (scale 1) and 1.6\,mas (scale 4). The SCC result does
not reveal any statistically different velocity components, nor does comparing
 SCC results individually on small- and large-scale structure.
Contrary to 3C\,273, the jet in 3C\,120 is not transversally resolved in the
MOJAVE images, which limit the possibility of detecting the flow stratification.
However, the high precision of the SCC measurement can be used in this case to
track the evolution of the flow speed with distance from the core.

\begin{figure}
    \centering
    \includegraphics{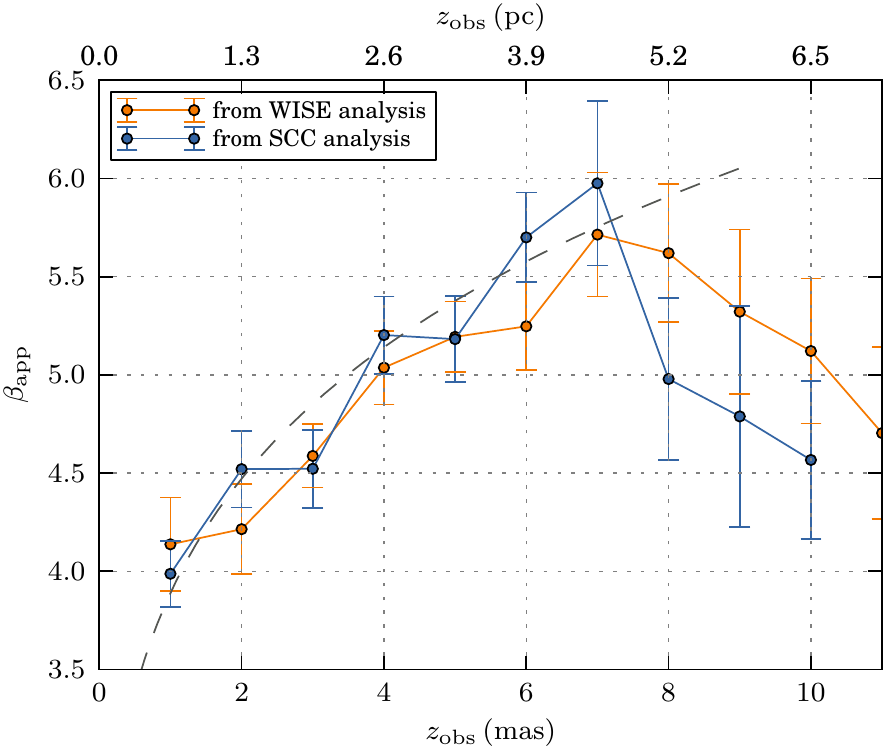}
    \caption
    {SCC analysis of the jet in 3C 120. The lack of transverse
      resolution in the MOJAVE images only warrants probing the
      longitudinal velocity distribution. Significant acceleration is
      detected up to a distance from the core of 7 mas, which is comparable to
      the
      acceleration detected from WISE analysis of the source.}
    \label{fig:3c120_acceleration}
\end{figure}

We computed the SCC individually for ten regions of the jet between a distance
from the core from 0.5 mas to 10.5 mas with a bin size of 1 mas. The result of
this analysis (Fig.~\ref{fig:3c120_acceleration}) reveals a significant flow acceleration from $\beta_
{\mathrm {app}} = 4.0\,\mathrm{c}
\pm 0.2$\,c at $z_{\mathrm{obs}} = 1$\,mas to $\beta_{\mathrm{app}} = 6.0\,\mathrm{c}\pm
0.4$\,c at $z_{\mathrm{obs}} = 7$\,mas, followed by an apparent deceleration of
the flow.

A similar analysis can be achieved using the results from the WISE analysis of
the source performed in~\cite{mertens_waveletbased_2015}. Apparent velocities of
each component detected in this study is obtained at several distances from the
core (DFC), between 1\,mas and 10\,mas with a step of 1\, mas, via linear fitting
of the component separation with time taking into account only the features 2
mas before and 2 mas after the DFC. An average apparent velocity is then
obtained for each DFC. The result is over-plotted in
Fig.~\ref{fig:3c120_acceleration} in orange, and we found it to be comparable to
the result obtained from the SCC analysis, proving the robustness of this last
method.

The observed apparent acceleration can be intrinsic flow acceleration or the
geometrical effect of a decreasing viewing angle due for example to the jet
bending. We estimate that for a Lorentz factor of 6, a decrease in viewing angle
of $\sim 15 \degree$ would be required to increase the apparent speed
from 4\,c to 6\,c. While the jet in 3C\,120 is known to display a significant
bending at large scales~\citep{benson_200_1984_mod}, only a slight jet direction
changes is observed at the scale of our analysis, and we can thus discard
apparent acceleration due to geometrical effects. There is also strong evidence
that flow acceleration can extend to a distance of few parsec from the
core~\citep{homan_mojave_2015}.

Using a viewing angle of $15
\degree$~\citep{agudo_recollimation_2012,hardee_modeling_2005}, the evolution of
the Lorentz factor with distance from the core can be described by a power-law
function $\gamma = a z^ {k}$ with $k \sim 0.3$. Magnetic jet
acceleration~\citep{vlahakis_theory_2015} or thermal/pressure induced
acceleration~\citep{georganopoulos_viewing_1998} would require an initial linear
acceleration phase with $\gamma \propto z$ for a conical jet like that in
3C\,120. The acceleration profile that we observe is hence more likely the
manifestation of a late-phase, slower, magnetic acceleration at which stage the
Poynting flux conversion starts to saturate~\citep{lyubarsky_asymptotic_2009}.

\section{Discussion}
\label{sc:discuss}

Stacked cross-correlation (SCC) introduced in this paper can be
applied to recover multiple velocity components from partially
overlapping emitting regions. The SCC extends and expands the
multi-scale cross-correlation (MCC) approach originally introduced for the
WISE analysis of astronomical images
\citep{mertens_waveletbased_2015}. While the MCC can partially recover
overlapping displacements in different SSP identified by WISE, the SCC
provides a fully statistical approach for robust identification of
multiple velocity components, which also includes  evaluation of statistical 
significance and estimation of uncertainty for each velocity component.

The SCC analysis of the transversely resolved jet in 3C\,273 reveals
two statistically significant velocity components corresponding to jet
plasma propagating at apparent superluminal speeds of $\sim 7.5\,c$
and $\sim 9\,c$. These two speeds may reflect the transverse
stratification of the flow. They may also be explained in a framework
of faster shock or plasma disturbances traveling through a slower
underlying flow.

In 3C\,120, the lack of transverse resolution in the MOJAVE data has
limited the SCC analysis to the longitudinal dimension of the flow. In
this setting, the SCC results have been found to be consistent with
the kinematic properties of the flow recovered previously using the
WISE analysis of the same data.

The method that we introduce  can be applied to sources with
composite structural evolution for which we have sufficient epochs of
observation, and the criteria is the ability to obtain velocity components with
statistically significant SCC coefficients. An ideal candidate for this analysis
would be the sub-parsec scale of the jet in M87 for which flow stratification
has already been suggested~\citep{mertens_longitudinal_2014}.

\begin{acknowledgements}
This research has made use of data from the MOJAVE database that is 
maintained by the MOJAVE team \citep{lister_mojave_2009}. FM was supported for this
research through a stipend from the International Max Planck Research School (IMPRS) for Astronomy and Astrophysics at the Universities of Bonn and Cologne.
\end{acknowledgements}

\bibliographystyle{aa}
\bibliography{biblio}

\end{document}